# Information Flow Coverage Metrics for Hardware Security Verification

Andres Meza
UC San Diego
anmeza@ucsd.edu

Ryan Kastner
UC San Diego
kastner@ucsd.edu

*Abstract*—Security graphs model attacks, defenses, mitigations, and vulnerabilities on computer networks and systems. With proper attributes, they provide security metrics using standard graph algorithms. A hyperflow graph is a register-transfer level (RTL) hardware security graph that facilitates security verification. A hyperflow graph models information flows and is annotated with attributes that allow security metrics to measure flow paths, flow conditions, and flow rates. Hyperflow graphs enable the understanding of hardware vulnerabilities related to confidentiality, integrity, and availability, as shown on the OpenTitan hardware root of trust under several threat models.

## I. Introduction

Security graphs have a long history of modeling attacks and defenses in computer and network systems [1]–[4]. Security graphs can be automatically generated from specifications and vulnerabilities [5]. Security-related attributes are added to the nodes and edges, enabling security metrics to understand and reason about an attack's severity and a vulnerability's effectiveness. The graph can be analyzed with efficient graph algorithms, e.g., shortest path, breadth- and depth-first search, cutsets, and min/max flows.

Hyperflow graphs translate the salient ideas of security graphs to RTL hardware security. Hyperflow graphs use attributes related to hardware information flow tracking (IFT) to model security vulnerabilities related to confidentiality, integrity, and availability. Hyperflow graphs enable hardware security metrics using graph algorithms. Hyperflow graphs are automatically built using open-source hardware compilers [6] and can be visualized with open-source tools [7].

Hyperflow analysis assesses information flows across the hardware, which is crucial for understanding security behaviors. While existing compiler dependency analysis provides information about data and control dependencies, it falls short of efficiently expressing and succinctly summarizing information flow relationships. Hyperflow analysis expands the traditional analysis to consider noninterference and other hyperproperties. The critical innovation is to augment the analysis to consider the security labels provided by hardware information flow tracking. This moves the analysis beyond trace properties to hyperproperties [8].

The hyperflow graph is annotated with simulation-based attributes to extract security-relevant information from a functional testbench. Thus, the hyperflow graph contains information relevant to functional simulation and data from the IFT security labels. Security label values summarize noninterference related to the assets in a design, which is crucial for verifying design properties pertaining to confidentiality, integrity, and integrity. Simulation data provide crucial attributes like the number of times a statement is executed and how many times a statement is executed with a specific security label.

The hyperflow graph naturally enables visualizations of information flows that help security verification engineers more readily understand a design's security posture. Property-driven hardware security verification [9] marks design assets and aims to understand where the information related to that asset could go and under what conditions. The hyperflow graph visualizes information flows by summarizing conditions and other relevant information related to the flows. This gives unique visual insights into the design's security and provides verification engineers with better intuition to assess weaknesses, understand vulnerabilities, and derive security properties.

The contributions of this work include the following:
- Introducing the hyperflow graph – a register transfer level security analysis and visualization model.
- Defining attributes and metrics on the hyperflow graph, which gives an understanding of hardware security vulnerabilities.
- Demonstrating the value of the hyperflow graph on the OpenTitan hardware root of trust.

## II. Background

### A. Graph-based Security Modeling and Metrics

Graphs are an important abstraction for modeling and understanding security attacks, defenses, vulnerabilities, and mitigations. Security graphs enable metrics to measure the system's vulnerabilities and mitigation effectiveness can be derived based on well-known and efficient graph algorithms. As such, there have been many security research efforts that use graphs to model attacks and defenses.

The privilege graph is one of the earliest works to use graphs to understand system security [1]. It models system access control where a node denotes a user (group) privileges, and edges represent the ability of one user to extend their privileges to another user (group). The privilege graph helps understand access control policies, define vulnerabilities via path search, and evaluate probabilities of successful attacks using attributes and graph search algorithms.

Attack trees formalize how to attack an asset [2], [3]. The tree's root is the attack's goal, and the leaf nodes define how to achieve that goal. The edges hold logical AND/OR labels that relate child nodes to their parents. The nodes have attributes that are Boolean or continuous values. Metrics can be defined on the tree to determine the best plans of attack, compare the costs of attacks, understand the potential vulnerabilities, and decide where to apply mitigations. Attack-defense trees more holistically model system security by adding countermeasures and describing their interactions with attacks [4].

Attack graphs are used extensively to analyze network security problems where nodes identify a logical statement of the network configuration and edges represent causal relationships between network configurations and the attacker's privileges [5], [10], [11]. The edges are annotated with attributes related to the probability of the attack success, time to perform the attack, or cost. Shortest path algorithms identify the attacks with the highest probability.

The attack graph literature is, by and large, focused on problems related to network security. However, some recent works adopt these ideas to issues more closely related to hardware security.

Koteshwara extends graph-based metrics to PCB-level server architectures [12]. Components and interfaces are nodes and edges,

respectively. They assign nodes with a base score – a metric of its risk severity based on the common vulnerability scoring systems (CVSS) where higher is more vulnerable. The base score is somewhat ad hoc, e.g., SPI, LPC, Processors, Ethernet sockets, VGA, USB, PCIe, UART, and I2C components are vulnerable (0.95) since they have Trojans listed in TrustHub. The root of trust is a lot less vulnerable (0.05) since it uses mitigations and protections. And the FLASH is not vulnerable (0.0) since the root of trust protects it. Saha et al. [13] make a preliminary attempt to extend this to system-on-chip architectures. Their analysis is performed on two small ($< 30$ vertices) SoC-level graphs using predefined impact probabilities.

Hyperflow graphs translate security-graph analysis to RTL hardware. Hyperflow graphs use hardware information flow tracking to model important information related to confidentiality, integrity, and availability. This enables succinct metrics that help understand potential weaknesses and determine vulnerabilities.

### B. Hardware Information Flow Tracking

Hardware information flow tracking (IFT) enables verification engineers to track the information stored in assets (e.g., registers, memory locations, and signals) as it propagates throughout the hardware. Hardware IFT automates security analysis with minimal effort. Using a property-driven design flow, verification engineers identify assets and describe where their information can and cannot propagate [9]. Hardware IFT facilitates the verification of hardware vulnerabilities related to confidentiality, integrity, availability, side channels, and other threat models [14].

Hardware IFT automates tracking assets using security labels. Every RTL variable is associated with a security label that denotes the presence of relevant information flows. Explicit flows occur as a result of direct data movement, i.e., RTL assignments. Implicit flows are context-dependent, i.e., a flow from a variable in a conditional statement to a variable in the assignments that can occur due to that conditional statement.

Hyperflow graphs are unique in their ability to model information flows on RTL designs, which provides a succinct summary of potential security vulnerabilities. Hyperflow graphs are annotated with attributes based on IFT-enhanced simulations. Graph algorithms applied to the hyperflow graph uncover weaknesses and provide metrics related to different vulnerabilities. We formally define hyperflow graphs in the next section, describe how to automatically derive a hyperflow graph from a RTL hardware design, and define its annotation process.

## III. HYPERFLOW GRAPH

A *hyperflow graph* is an intermediate representation (IR) that aids security analysis and visualization. Hyperflow graphs depict the relationships between hardware design variables, including direct flows, conditional flows, information relevant to security verification (assets, boundaries), functional simulation (assignments executed), and other important information required for hardware security analysis.

### A. Formal Definition

A hyperflow graph $G$ is a directed graph $G = (V, E)$ where:
- $V$ is a set of vertices annotated with attributes;
- $E$ is a set of directed edges annotated with attributes.

Each vertex $v \in V$ is an ordered pair $(\mathtt{s}, \mathtt{M}^v)$ with $\mathtt{s}$ being a single design signal in the RTL design under analysis and $\mathtt{M}^v$ being a set containing simulation metadata associated with the signal $\mathtt{s}$.

$$\begin{aligned} V &= \{\quad v_1, \quad v_2, \quad ..., \quad v_{n-1}, \quad v_n \quad \} \\ &= \{(\mathtt{s}_1, \mathtt{M}_1^v), (\mathtt{s}_2, \mathtt{M}_2^v), ..., (\mathtt{s}_{n-1}, \mathtt{M}_{n-1}^v), (\mathtt{s}_n, \mathtt{M}_n^v)\} \end{aligned}$$

Each element of a *vertex-metadata* set $\mathtt{M}^v$ is an ordered triple of the form $(\mathtt{t}, \mathtt{val}, \mathtt{val}^\tau)$ where
- $\mathtt{t}$ is a specific time in the simulation
- $\mathtt{val}$ is the functional value of signal $\mathtt{s}$ at time $\mathtt{t}$
- $\mathtt{val}^\tau$ is the taint value (i.e., security label value) of signal $\mathtt{s}$ at time $\mathtt{t}$.

Each edge $e \in E$ is an ordered triple $(\mathtt{s}^t, \mathtt{s}^h, \mathtt{M}^e)$ where $\mathtt{s}^t$ and $\mathtt{s}^h$ are the signals representing the tail and head of edge $e$, respectively. $\mathtt{M}^e$ is a set containing simulation metadata associated with the information flow from signal $\mathtt{s}^t$ to signal $\mathtt{s}^h$. The hyperflow graph can model implicit, explicit, control, data, and timing information flows.

$$\begin{aligned} E &= \{\quad e_1, \quad ..., \quad e_m \quad \} \\ &= \{(\mathtt{s}_1^t, \mathtt{s}_1^h, \mathtt{M}_1^e), ..., (\mathtt{s}_m^t, \mathtt{s}_m^h, \mathtt{M}_m^e)\} \end{aligned}$$

Each element of an *edge-metadata* set $\mathtt{M}^e$ is an ordered pair of the form $(\mathtt{t}^s, \mathtt{t}^e)$ where
- $\mathtt{t}^s$ is a simulation time at which all conditions necessary for the information flow modeled by this edge to occur are satisfied
- $\mathtt{t}^e$ is a simulation time (following $\mathtt{t}^s$) at which one or more conditions necessary for the information flow modeled by this edge to occur are no longer satisfied.

### B. Characteristics of Hyperflow Edges & Flows

Hyperflow graph edges represent a *potential* information flow. The information flows are derived from assignments in the RTL design. More precisely, an edge $(v_\mathtt{X}, v_\mathtt{Y})$ corresponds to an assignment in the RTL code where signal $\mathtt{Y}$ is on the assignment's LHS and signal $\mathtt{X}$ is (1) an operand on the assignment's RHS, (2) an operand in the assignment's guarding conditions, or (3) both (1) and (2).[1] Case (1) is an *explicit flow* from $\mathtt{X}$ to $\mathtt{Y}$ and case (2) is an *implicit flow*.

Information flows can also be characterized based on their conditional nature. A *conditional flow* has predicates (or guarding conditions) that must be satisfied for it to occur. Conversely, an *unconditional flow* has no predicates and occurs continuously. Conditional flow predicates are the *cumulative* predicates, i.e., they include all of the nested guarding conditions surrounding the construct from which the flow originated.

### C. Automated Hyperflow Graph Generation

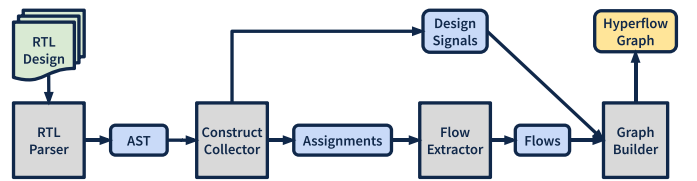

Fig. 1: The automated process of constructing a hyperflow graph from the source RTL code of a hardware design.

Hyperflow graphs are constructed via the automated process shown in Figure 1. The input is the RTL code (e.g., Verilog files) and the output is a hyperflow graph for that hardware design.

The graph construction process has four major steps. The first step parses the input RTL code to generate an abstract syntax tree (AST). We utilize SystemVerilog as the HDL and slang [6] for the parser. The general ideas translate to other HDLs and parsers. The second step traverses the AST to identify the AST nodes/subtrees corresponding to signals or information flows between signals. Signals are extracted from SystemVerilog declaration constructs including `reg`, `wire`, and `logic`. Similarly, flows are extracted from SystemVerilog assignment constructs including `continuous_assign`, `blocking_assignment`, and `nonblocking_assignments`, and SystemVerilog constraint/conditional constructs including `if`, `else`, or `case`. The third step extracts the information flows from constructs collected in the previous step. As mentioned in Section III-B, flows are extracted from the assignments and the guarding conditions of those assignments. The final step turns the extracted design signals and information flows into the vertices and edges, respectively, of a hyperflow graph as defined in Section III.

---

[1]When a signal $\mathtt{X}$ is an operand on the assignment's RHS, and in the assignment's guarding conditions, each flow is represented as a unique edge.

### D. Hyperflow Graph Attributes and Annotation Process

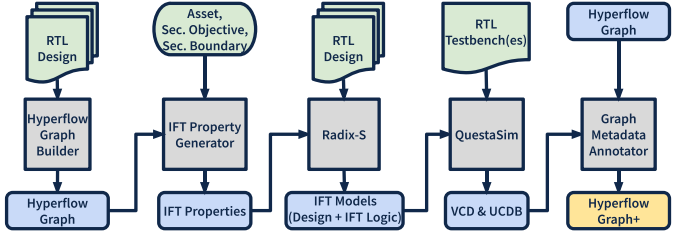

Fig. 2: The automated process of annotating a hardware design's hyperflow graph with simulation metadata including, but not limited to, the functional values and IFT security labels of design signals.

Hyperflow graphs can be annotated with various forms of metadata in order to derive insights that could be not derived by the unannotated hyperflow graph alone. In this paper, hyperflow graphs are annotated with simulation metadata via the automated process shown in figure 2. The input to the automated process is (1) the RTL code of a particular hardware design, (2) a list of design assets along with their security objectives and security boundaries, and (3) at least one RTL testbench for the given hardware design. The output of the annotation process is a hyperflow graph whose vertices have been annotated with their respective design signal's functional and taint values and whose edges have been annotated with their respective flow's activation times. The graph annotation process can be broken into 5 major steps.

The first step of annotation process creates an unannotated hyperflow graph (see Section III-C). This step is only necessary if a hyperflow graph for the current hardware design does not already exist.

The second step of the process utilizes a hyperflow graph and a list of design assets along with their security objectives and security boundaries in order to generate a set of IFT properties. These IFT properties will be used to generate the Radix security monitors (i.e., IFT models generated by Radix-S) that will be simulated alongside the original design in later steps. An IFT property for a given triple of (asset, objective, boundary) will look like one of the following IFT properties.

```
assert iflow(
  asset =/=> {signal(s) outside sec. boundary}
);

assert iflow(
  {signal(s) outside sec. boundary} =/=> asset
);
```

The first IFT property verifies the confidentiality of the asset by specifying that information from the asset should not leak to the set of signals outside of its security boundary. The second property verifies integrity of the asset by specifying that information from signals outside of the asset's security boundary should not leak to asset.

The third step uses Cycuity's Radix-S tool in order to automatically generate the Radix security monitors (i.e., IFT models) needed to generate/track the security labels during simulation. Radix-S constructs a security monitor for every IFT property. Each security monitor tracks information leakage from all source signals (i.e., the signal(s) on the left hand side of the no-flow operator) to all other design signals.

The fourth step uses a commercial simulator (i.e., Questa Advanced Simulator) in order to simulate the original design and all security monitors with the provided testbench(es). During simulation, traces of the original design and the security monitors are stored for later use. The traces of the original design contain the functional values for all design signals whereas the traces of the security monitors contain the taint values for all design signals.

The final step in the process parses the stored traces (i.e., vcd files) and maps the simulation metadata onto the appropriate vertices and edges.

## IV. INFORMATION FLOW METRICS

In this section, we introduce a set of quantitative information flow coverage metrics that can be derived from the hyperflow graph. For each coverage metric, we describe how it is derived and what insight it provides. It should be noted that there are other useful metrics that could be derived from the hyperflow graph which are not included in this section for the sake of brevity and clarity.

### A. Metric #1: Signal Connectivity

The signal connectivity metric (SCM) is a Boolean metric which indicates if it is possible for information to flow from some signal A to some other signal B under the normal execution of design. If the hyperflow graph does not contain any paths from signal A to signal B, then there is no way for information to flow from signal A to signal B. However, if there is at least one path from signal A to signal B, further analysis is needed to determine if the path is realizable. If a path from signal A to signal B only utilizes unconditional flows/edges then it is possible for information to leak from A to B via this path. However, if a path from signal A to signal B utilizes any conditional edges then one must perform reachability analysis to determine if the sequence of design states needed to activate each edge/flow from A to B is actually realizable based on the logic/FSMs of the rest of the design. An alternative to performing this formal reachability analysis is to perform IFT simulations of the design where signal A is the IFT source (i.e., the signal whose information is being marked with security labels). If the security label for B changes from 0 to 1 at any point during the simulation then it is possible for information to flow.

### B. Metric #2: Path Activation

The path activation metric (PAM) quantifies how many of the paths from signal A to signal B were activated during a particular simulation. A path is considered to be activated when information from signal A flows to signal B in the manner described by the path's vertices and edges.

$$\text{PAM}(G, A, B) = \frac{\text{\# of Activated Paths}}{\text{Total \# of Paths from } A \text{ to } B}$$

There are two major steps to deriving PAM from an annotated hyperflow graph. The first major step in the process collects the set of times when information from signal A reached signal B.

This is done by examining the security labels stored in signal B's respective vertex metadata set. The second major step in the process determines, for every time collected in the previous step, which path enabled information from signal A to reach signal B. This is done by reversing the edges of the hyperflow graph and then performing a constrained path search from signal B to signal A. The constraint is placed on which edges can be taken during the search. Specifically, an edge can only be taken if the flow it represents is the flow which led to information from A reaching the current vertex/signal at a particular time $t$. This can be determined by finding the edge or edges whose constraints are satisfied at $t$ and whose source vertices have a corresponding non-zero security label at time $t$ as well.

*C. Metric #3: Signal Proximity*

The signal proximity metric (SPM) indicates how close information from some signal A is to reaching some signal B. The notion of "close-ness" or distance is measured in hyperflow graph edges. For a given simulation time t, SPM will return the smallest number of edges separating signal A's information from signal B.

$$\text{SPM}(G, A, B) = \text{min. \# of edges between } A\text{'s info. and } B$$

The first step checks signal B's security label metadata to determine if information from signal A has already reached signal B. If information has already reached signal B SPM is 0. Otherwise, the process continues. The second step collects the set of vertices/signals which contain information from signal A at time t. This can be done by performing a BFS from signal A until signal B is reached. While performing the BFS, each vertex/signal's security label metadata is checked to see if it contains information from signal A. The final step computes the shortest path between all of the candidate signals and signal B, and then returns the length (i.e., number of edges) of the shortest path.

*D. Metric #4: Local Information Flow Rate*

sThe local information flow rate (LIFR) metric quantifies how much information from signal A is flowing into signal B within a given time window. The amount of information flow is measured in bits.

$$\text{LIFR}(G, B, t_1, t_2) = \frac{\text{SLBT}(B, t_1, t_2)}{t_2 - t_1}$$

The $\text{SLBT}(B, t_1, t_2)$ term returns the number of times any bit in B's security label transitioned from 0 to 1 with the given time window.

*E. Metric #5: Global Information Flow Rate*

The global information flow rate (GIFR) metric determines how much information from signal A is flowing throughout all into signal B within a given time window.

## V. EXPERIMENTS

In this section we briefly introduce the OpenTitan hardware root-of-trust and then describe the information flow metrics derived on its one-time programmable (OTP) memory controller via its annotated hyperflow graph.

*A. OpenTitan*

OpenTitan is a commercial-grade, open-source hardware root-of-trust [15]. In the semiconductor industry, hardware roots-of-trust are used to perform a wide range of security-critical tasks such as secure boot, management and configuration of operating modes (e.g., normal vs debug), and management of sensitive data (e.g., cryptographic keys). Designed to support a wide range of these hardware security use cases, OpenTitan features a security-enhanced RV32IMCB RISC-V Ibex core, various security peripherals (e.g., AES, KMAC, HMAC), multiple memories (e.g., ROM, eFLASH, SRAM, OTP) with dedicated controllers for access control and scrambling purposes, and various IO peripherals. We showcase the utility of the hyperflow graph and associated metrics on the one-time programmable (OTP) memory controller in this experiment.

The OTP controller is a non-volatile memory that stores keys and other security assets. Thus, its security validation is crucial for the correct and secure operation of OpenTitan. Within the OTP controller exists the `RndCnstKey` asset – a netlist constant used as a key encryption key for security assets stored in the OTP memory. FLASH and SRAM scrambling keys, device root secret and unlock tokens, and other assets stored in the OTP memory are encrypted with `RndCnstKey` so that these assets are not stored in plaintext in the OTP. Thus, it is critical for `RndCnstKey` to remain confidential; access to `RndCnstKey` would enable an adversary to decrypt security-critical data from the OTP memory. Any information from `RndCnstKey` should remain within the OTP controller; no knowledge about `RndCnstKey` should leak outside the controller. To verify this, we perform an IFT-instrumented simulation of OpenTitan which tracks information from `RndCnstKey` to any signal in the OpenTitan.

*B. Hyperflow Graph*

Figure 3 shows various visualizations of the hyperflow graph of the OTP controller. In total, there are 11,579 vertices (i.e., registers) and 32,084 edges (i.e., potential information flows) in the hyperflow graph. Figure 3a shows all of the explicit edges and implicit edges of the hyperflow graph. Figure 3b shows only the explicit edges whereas 3c shows only the implicit edges. Lastly, 3d shows one of the paths which enables information from `RndCnstKey` to flow outside of the controller. The creation of such visualizations is made possible by the metadata/attributes attached to the vertices and edges of the hyperflow graph. This hyperflow graph is used to derive the metrics reported in Table I.

*C. IFT Coverage Metrics*

Following the IFT-instrumented simulation of OpenTitan, we annotate the OTP controller's hyperflow graph with security labels and other simulation metadata as described in SectionIII-D. We then derive the four proposed security metrics for all of the outputs of the OTP controller. Table I reports the derived metrics.

## VI. RELATED WORKS

*A. Hardware Security Metrics*

Power side channels have some of the most effective and widely-used metrics in hardware security. They use statistical metrics [16]–[19], which have a direct mapping to that threat model. Unfortunately, these metrics do not effectively map to more general hardware security threat models related to confidentiality, integrity, and availability that we consider.

Rostami et al. summarize metrics for Trojans, IP piracy, reverse engineering, side channels, and counterfeiting [20], which represent different threat models than proposed here. DSeRC has two metrics related to fault injection and Trojans [21].

Various hardware IFT verification tools provide binary decisions (two-value logic metrics) about the violation of properties [22]–[24]. These are useful for detecting the existence of potential vulnerabilities, but less valuable for understanding the vulnerabilities.

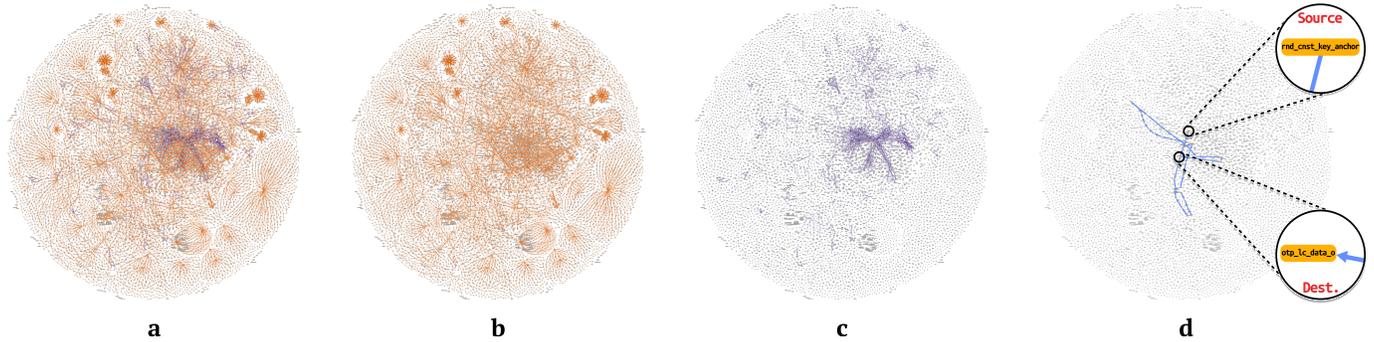

a  b  c  d

Fig. 3: Part a) shows the hyperflow graph for the OTP controller which contains 11,579 vertices (i.e., registers) and 32,084 edges (i.e., potential information flows). Of these 32,084 edges, 27,218 are explicit flows (shown in Part b)) and 4,866 are implicit flows (shown in Part c)). Part d) shows the path through which information from the OTP scrambler asset `RndCnstKey` leaked to the output of the OTP controller (`otp_lc_data_o.test_unlock_token`) during an IFT simulation. Interactive versions of these hyperflow graphs are accessible at hyperflowgraphs.com for enhanced viewing and exploration.

TABLE I: This table reports various information flow coverage metrics on the outputs of the OpenTitan's OTP memory controller with respect to information originating from RndCnstKey. "SCM" indicates the signal connectivity metric from RndCnstKey to all listed outputs. "PAM*" indicates the path activation metric for the paths from RndCnstKey to all listed outputs. "SPM" indicates the signal proximity metric for RndCnstKey's information and all outputs. "LIFR (x%)" indicates the local information flow rate for all outputs after 25%, 50%, 75%, and 100% of IFT-instrumented simulation has completed. The simulation was driven by the OpenTitan's "chip_sw_otp_ctrl_smoketest" testbench.

| No. | Output Ports of OTP Memory Controller | SCM | PAM* | SPM | LIFR (25%) | LIFR (50%) | LIFR (75%) | LIFR (100%) |
|---|---|---|---|---|---|---|---|---|
| 1 | /ot_top/u_otp_ctrl/cio_test_en_o | FALSE | 0 | ∞ | 0 | 0 | 0 | 0 |
| 2 | /ot_top/u_otp_ctrl/cio_test_o | FALSE | 0 | ∞ | 0 | 0 | 0 | 0 |
| 3 | /ot_top/u_otp_ctrl/intr_otp_error_o | TRUE | 0 | 26 | 0 | 0 | 0 | 0 |
| 4 | /ot_top/u_otp_ctrl/intr_otp_operation_done_o | TRUE | 0 | 26 | 0 | 0 | 0 | 0 |
| 5 | /ot_top/u_otp_ctrl/lc_otp_vendor_test_o | FALSE | 0 | ∞ | 0 | 0 | 0 | 0 |
| 6 | /ot_top/u_otp_ctrl/otp_ast_pwr_seq_o | FALSE | 0 | ∞ | 0 | 0 | 0 | 0 |
| 7 | /ot_top/u_otp_ctrl/otp_obs_o | FALSE | 0 | ∞ | 0 | 0 | 0 | 0 |
| 8 | /ot_top/u_otp_ctrl/prim_tl_o | FALSE | 0 | ∞ | 0 | 0 | 0 | 0 |
| 9 | /ot_top/u_otp_ctrl/edn_o | TRUE | 1 | 0 | 0 | 1 | 1 | 1 |
| 10 | /ot_top/u_otp_ctrl/lc_otp_program_o | TRUE | 1 | 0 | 0 | 2 | 2 | 2 |
| 11 | /ot_top/u_otp_ctrl/pwr_otp_o | TRUE | 1 | 0 | 0 | 2 | 2 | 2 |
| 12 | /ot_top/u_otp_ctrl/alert_tx_o | TRUE | 1 | 0 | 0 | 4 | 4 | 6 |
| 13 | /ot_top/u_otp_ctrl/core_tl_o | TRUE | 1 | 0 | 0 | 0 | 0 | 135 |
| 14 | /ot_top/u_otp_ctrl/otbn_otp_key_o | TRUE | 1 | 0 | 0 | 128 | 128 | 193 |
| 15 | /ot_top/u_otp_ctrl/flash_otp_key_o | TRUE | 1 | 0 | 0 | 128 | 128 | 259 |
| 16 | /ot_top/u_otp_ctrl/otp_keymgr_key_o | TRUE | 1 | 0 | 0 | 513 | 513 | 513 |
| 17 | /ot_top/u_otp_ctrl/otp_hw_cfg_o | TRUE | 1 | 0 | 0 | 644 | 644 | 644 |
| 18 | /ot_top/u_otp_ctrl/sram_otp_key_o | TRUE | 1 | 0 | 0 | 384 | 384 | 771 |
| 19 | /ot_top/u_otp_ctrl/otp_lc_data_o | TRUE | 1 | 0 | 0 | 1102 | 1102 | 1102 |

Contreras et al. develop a vulnerability analysis specifically for test structures with experiments related to cryptographic cores, Trojans, and access control [25]. Their analysis performs a path traversal at the gate level and calculates a distance – the number of gates from the asset to control or observation points. They argue that longer paths are more secure. Hyperflow graphs work at the RTL which provides additional information related to control structures and use noninterference to more accurately understand information flows.

Trust coverage is a quantitative metric for the trust level of a third party IP [26]. It includes three vector-based coverage metrics related to function, structure, and assets. Asset coverage is the most relevant metric that can be computed upon the hyperflow graph. Asset coverage measures the number of observable outputs that a specified asset can reach. They evaluate their metrics for hardware Trojans.

Quantitative information flow characterizes the information leakage magnitude [27]. There are attempts to translate this to hardware [28]–[30] and focus on cryptographic cores and Trojans. However, it is challenging to translate quantitative IFT outside of these threat models.

SCRIPT is a framework to evaluate side channel leakage using vulnerability metric based on information flow tracking [31]. SCRIPT uses a gate-level path sensitization formal method to identify registers where information can flow. Formal methods suffer from the ability to scale; their techniques resort to heuristics, e.g., to avoid full-sequential ATPG, which they state is "ineffective and time-consuming" [25]. Furthermore, they target power side channel. Our techniques subsume this work in scope and scalability by using simulation-based IFT analysis. Our techniques could be applied to their IFT-based analysis to find target registers more efficiently.

HW2VEC is an automated tool for converting a hardware design to a Euclidian graph embedding that is subsequently used as feature input to a graph neural network [32]. They show how to use this graph transformation to train a binary neural network that indicates whether the design has a hardware Trojans. The other example develops a neural network to detect if one design is pirated from another. They solve a different problem than what we propose.

We are not aware of any work that uses metrics based upon simulation-based IFT. The benefits of our technique include: 1)

applying noninterference rather than just direct affectability, 2) understanding testbench effects on the security and vice-versa, and 3) providing a measure of flow rates with respect to implicit and explicit flows.

## VII. Conclusion

In this paper, we introduced the hyperflow graph, a register transfer level security analysis and visualization model, and defined attributes and metrics on the hyperflow graph, which gives an understanding of hardware security vulnerabilities. We demonstrated the value of the hyperflow graph on the OpenTitan hardware root of trust by using the proposed hyperflow graph and metrics to both visualize and understand the flow of information from a security-critical asset in OpenTitan's OTP memory controller.